# ON THE INTERPLAY BETWEEN $T_c$-INHOMOGENEITIES AT LONG LENGTH SCALES AND THERMAL FLUCTUATIONS AROUND THE AVERAGE SUPERCONDUCTING TRANSITION IN CUPRATES

FÉLIX VIDAL, JOSÉ ANTONIO VEIRA, JESÚS MAZA, JESÚS MOSQUEIRA AND CARLOS CARBALLEIRA
*Laboratorio de Bajas Temperaturas y Superconductividad, Fac. de Física, U. Santiago de Compostela, E-15706 Spain*

ABSTRACT. We review at an introductory and pedagogical level some aspects of the interplay between inhomogeneities at long length scales (at length scales much bigger than any characteristic length for superconductivity, in particular than the superconducting coherence length amplitude, $\xi(T)$, even at temperatures relatively close to $T_c$) and the intrinsic fluctuations of Cooper pairs above $T_c$ in high temperature cuprate superconductors (HTSC). These inhomogeneities at long length scales do not directly affect the thermal fluctuations, but they may deeply affect, together (and entangled!) with the thermal fluctuations, the measured behaviour of any observable around the transition. The emphasis is centered on the role played by the presence of $T_c$-inhomogeneities, as those associated with oxygen content inhomogeneities, at these long length scales and uniformly or non-uniformly distributed in the samples, on the in-plane transport properties in inhomogeneous HTSC crystals. For completeness, we will also summarize some results on this interplay when various types of inhomogeneities (i.e., structural and stoichiometric, uniformly and non-uniformly distributed) may be simultaneously present.

## Contents

**1. Introduction**

**2. $T_c$-inhomogeneities uniformly distributed**
    2.1. $T_c$-inhomogeneity effects on the in-plane magnetoconductivity above the average transition temperature.
    2.2. Splitting of the resistive transition: Intrinsic double superconducting transitions versus extrinsic effects.

# 1. Introduction

The interplay between the intrinsic thermal fluctuation effects around the superconducting transition and the extrinsic inhomogeneity effects associated with stoichiometric and structural inhomogeneities at different length scales, was already an important problem in low temperature superconductors (LTSC). For instance, in summarizing in 1978 the effects of the thermal fluctuations of Cooper pairs on the electrical resistivity, $\rho(T)$, above the superconducting transition in metallic films, effects that had been actively studied in the last ten years, Kosterlitz and Thouless concluded that the onset of the observed rounding of $\rho(T)$ "*may alternatively be a result of film inhomogeneities*". [1] In high temperature superconducting cuprates (HTSC), the dilemma between sample inhomogeneities and thermodynamic fluctuations above $T_C$ was earlier stated by Bednorz and Müller in their seminal work, [2] although they formulated the alternative in an opposite way to that done by Kosterlitz and Thouless for LTSC´s: After having indicated that the observed rounding of $\rho(T)$ around $T_C$ in their LaSCO compounds may be due to inhomogeneities, Bednorz and Müller concluded that "*the onset (of the $\rho(T)$ drop) can also be due to fluctuations in the superconducting wave functions*". In fact, mainly due to the smallness of the superconducting coherence length amplitudes (at 0 K), $\xi(0)$, which are anisotropic but in all directions of the order of the interatomic distances, both effects, those associated with the intrinsic thermal fluctuations and those with the extrinsic inhomogeneities, may be very important in the HTSC. This is mainly due, in the case of the thermal fluctuations, to the fact that a small $\xi(T)$ leads to a small coherent volume, which will contain very few strongly correlated Cooper pairs. These fluctuation effects are also enhanced by the layered nature of the HTSC, which may lead these materials to behave as quasi bi-dimensional superconductors (still reducing, then, their superconducting coherent volume) and their high $T_C$, which increase then the available (gratis!) thermal agitation energy, of the order of $k_B T_C$ (where $k_B$ is the Boltzmann constant) around their superconducting transition. [3]

In the case of the inhomogeneities, the smallness of $\xi(T)$ makes the different superconducting properties of these materials very sensitive to the presence of inhomogeneities, even when they have very small characteristic length, of the order of $\xi(T)$. In addition, their layered nature and the complexity of their chemistry enhance the probability of the presence of extrinsic inhomogeneity effects in real HTSC compounds. When they are present at long length scales (i.e., at length scales much bigger than any characteristic length in the system, as the magnetic field penetration length or, mainly, the superconducting coherence length, $\xi(T)$, even for temperatures relatively close to $T_C$), these inhomogeneities will not *directly* affect the thermal fluctuations themselves, but still they may deeply affect, together with the thermal fluctuations, the measured behaviour of any observable around the superconducting transition. [3]

In this paper, we will summarize some of our results on the interplay between the inhomogeneities of $T_C$ at long length scales and the thermal fluctuations around the average superconducting transition. In particular, we will indicate through some examples how to disentangle the intrinsic effects (associated with the thermal fluctuations) from the extrinsic ones (associated with inhomogeneities). But, in any case these examples are aimed also to illustrate the fact that in analyzing an anomalous critical behaviour of any observable it is important to carefully check the possible presence of extrinsic inhomogeneity effects.

## 2. $T_c$-inhomogeneities uniformly distributed

Probably one of the most common types of inhomogeneities in HTSC are the critical temperature ($T_C$) inhomogeneities at long length scales. These inhomogeneities may be produced by, for example, oxygen content inhomogeneities at these long length scales. But, in addition to stoichiometric inhomogeneities, there are other possible causes for $T_C$-inhomogeneities in LTSC and HTSC compounds, such as local strains, [4] or low dimensionality effects. [5] In fact, these last effects may appear when, for instance, the sample dimension in one direction is smaller than the coherence length amplitude in that direction (so, in this case, the inhomogeneities will be in the so called small-length-scale regime, and the fluctuations will be also directly affected). [5]

An expected but non-trivial effect of these $T_C$-inhomogeneities at long length scales and uniformly distributed in the samples is that they round the critical behaviour of different observables around the superconducting transition, in competition with the intrinsic rounding effects associated with thermal fluctuations. [6] In the first part of this

Section, we summarize some of our results on this type of $T_c$-inhomogeneity effects on the in-plane resistivity and on the magnetoresistivity in some inhomogeneous HTSC crystals. Also, we will indicate how to disentangle the intrinsic and the extrinsic effects, around the average $T_c$, on both observables. In the second part of this subsection, we will summarize some of our results on the apparent double resistivity transition, effects that may be easily explained in terms of inhomogeneities. [7]

## 2.1. $T_c$-INHOMOGENEITY EFFECTS ON THE MAGNETOCONDUCTIVITY ABOVE THE AVERAGE TRANSITION TEMPERATURE

An illustrative example of the interplay between the intrinsic fluctuation effects and the extrinsic effects associated with inhomogeneities of characteristic lengths much bigger than the superconducting coherence length, $\xi(T)$, in all directions, is provided by the in-plane magnetoconductivity, $\sigma_{ab}(T,H)$, in presence of $T_c$-inhomogeneities at these long length scales and *uniformly distributed* in the sample. In the case of the electrical conductivity (in absence of applied magnetic field) such an interplay was first studied by Maza and Vidal (MV)[6] by using an effective medium approach. These results were recently extended by Pomar et al. [7] to study the influence of the $T_c$ inhomogeneities on the in-plane magnetoconductivity in inhomogeneous $Bi_2Sr_2Ca_1Cu_2O_8$ (Bi-2212) crystals. We will summarize here these last results.

The effective (measured) in-plane magnetoconductivity $\sigma^e_{ab}(T,H)$, may be related to the intrinsic one (the conductivity measured in an ideal, homogeneous, crystal), $\sigma_{ab}(T,H)$, through the MV expression, based on the Bruggeman effective medium approach, [6,8]

$$\int_0^\infty \frac{\sigma_{ab}(T,H) - \sigma^e_{ab}(T,H)}{\sigma_{ab}(T,H) + 2\sigma^e_{ab}(T,H)} Q(\sigma_{ab},T) d\sigma_{ab} = 0 \qquad (1)$$

Here, $Q(\sigma_{ab},T)d\sigma_{ab}$ is the local conductivity distribution, i.e., the volume fraction of the sample with a local (or intrinsic) conductivity between $\sigma_{ab}(T,H)$ and $\sigma_{ab}(T,H) + d\sigma_{ab}(T,H)$. [6,7] This local conductivity may be written as the sum of the normal conductivity plus the corrections due to the thermal fluctuations,

$$\sigma_{ab}(T,H) = \sigma_{abB}(T,H) + \Delta\sigma_{ab}(T,0) + \Delta\tilde{\sigma}_{ab}(T,H), \qquad (2)$$

where here $\Delta\tilde{\sigma}_{ab}(T,H) \equiv \sigma_{ab}(T,H) - \sigma_{ab}(T,0)$.

To approximate $Q(\sigma_{ab},T)$ in Eq. (1), we may note first that the basic effects of the stoichiometric inhomogeneities on $\sigma^e_{ab}(T,H)$, and therefore on $Q(\sigma_{ab},T)$, are due to the associated critical temperature inhomogeneities. For the corresponding distribution of $T_{c0}$'s, we will follow the MV procedure, which assumes a spatial Gaussian distribution characterized by the mean value of the critical temperature, $\bar{T}_{c0}$, and by the standard deviation $\Delta\bar{T}_{c0}$. Thus, the conductivity distribution may be written as

$$Q(\sigma_{ab},T)d\sigma_{ab} = \frac{2}{\sqrt{\pi}\Delta \overline{T}_{c0}} \exp\left\{-\left(\frac{T_{c0}-\overline{T}_{c0}}{\Delta \overline{T}_{c0}}\right)^2\right\}dT_{c0} \qquad (3)$$

As the exponential function in Eq. (3) is rapidly decreasing, to evaluate the integral in Eq. (1) we may change the integration limit to the interval $\overline{T}_{c0} \pm 2\Delta \overline{T}_{c0}$. Then, Eq. (1) becomes

$$\int_{\overline{T}_{c0}-2\Delta\overline{T}_{c0}}^{\overline{T}_{c0}+2\Delta\overline{T}_{c0}} \frac{\sigma_{ab}(T,H)-\sigma_{ab}^e(T,H)}{\sigma_{ab}(T,H)+2\sigma_{ab}^e(T,H)} \frac{C}{\Delta\overline{T}_{c0}} \exp\left\{-\left(\frac{T_{c0}-\overline{T}_{c0}}{\Delta\overline{T}_{c0}}\right)\right\}dT_{c0} = 0 \qquad (4)$$

where $C = 0.5448$ arises from the normalization conditions. This expression links, therefore, the intrinsic and the effective conductivities through $\Delta\overline{T}_{c0}$, the standard deviation of critical temperatures due to inhomogeneities. The corresponding *effective* in-plane paraconductivity and fluctuation induced in-plane magnetoconductivity may be defined by just using $\sigma_{ab}^e(\overline{\varepsilon},H)$, calculated through Eq. (4), in the conventional definitions of the in-plane paraconductivity, $\Delta\sigma_{ab}(\varepsilon)$, and fluctuation induced magnetoconductivity, $\Delta\tilde{\sigma}_{ab}(\varepsilon)$, i.e.,

$$\Delta\sigma_{ab}^e(\overline{\varepsilon},0) \equiv \sigma_{ab}^e(\overline{\varepsilon},0) - \sigma_{abB}^e(\overline{\varepsilon},0), \qquad (5)$$

and

$$\Delta\tilde{\sigma}_{ab}^e(\overline{\varepsilon},H) \equiv \sigma_{ab}^e(\overline{\varepsilon},H) - \sigma_{abB}^e(\overline{\varepsilon},0), \qquad (6)$$

where $\overline{\varepsilon} \equiv (T-\overline{T}_{c0})/\overline{T}_{c0}$ is the reduced temperature associated to the average transition temperature, $\overline{T}_{c0}$. Let us note here that, as already stressed in Ref. 6, the $T_c$-inhomogeneities will affect more strongly the observables with stronger intrinsic $\varepsilon$-dependencies. In the Bi-2212 crystals, where the thermal fluctuations of Cooper pairs around $T_c$ are strongly two-dimensional (2D), the intrinsic critical exponents in the mean field region above $T_{c0}$ are at around -3 and -1 for, respectively, $\Delta\tilde{\sigma}_{ab}(T,H)$ (in the weak $H$-limit) and $\Delta\sigma_{ab}(\varepsilon)$.[7] So, we may expect that $\Delta\tilde{\sigma}_{ab}(\varepsilon)$ is going to be much more affected by inhomogeneities than $\Delta\sigma_{ab}(\varepsilon)$.

As an example of application of the above approach, in Fig. 1(a) we present the in-plane paraconductivity (open circles) and of the fluctuation induced in-plane magnetoconductivity (open triangles and squares) measured by Pomar et al. [7] in a Bi-2212 crystal. The lines in this figure correspond to the intrinsic (without inhomogeneities) direct order parameter fluctuation effects calculated in a layered material with two layers per periodicity length and with the same Josephson coupling strength between adjacent layers, which is the case well suited for Bi-2212 crystals (see Refs. 7 and 9). The value of the in-plane superconducting coherence length amplitude used here was $\xi_{ab}(0) \approx 1$ nm, which agrees with the one found by analyzing other fluctuation effects in this system (see Ref. 7). The results of Fig. 1 (a) show that whereas for the paraconductivity the agreement between the theory and the experimental data is excellent, there is a dramatic disagreement between the measured and the calculated fluctuation induced in-plane magnetoconductivity. Such a disagreement is confirmed by the results presented in Fig. 2 (a), where the measured excess conductivity is scaled as a function of $(T-T_c(H))(TH)^{-1/2}$, which is expected to hold for two-dimensional thermal fluctuations. [10] As it can be seen in this figure, there is no scaling below $T_{c0}$. These results may easily be understood at a qualitative level in terms of $T_c$-inhomogeneities, on the grounds of the comments presented at the end of the above paragraph.

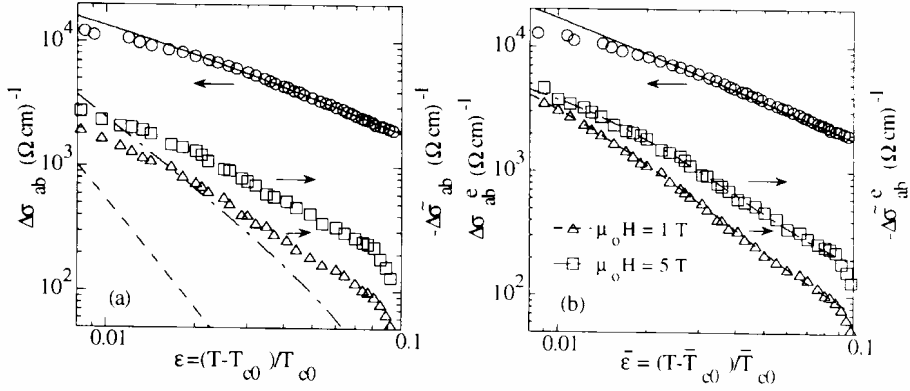

*Figure 1*. Reduced temperature dependence of the measured in-plane paraconductivity and of measured in-plane fluctuation induced magnetoconductivity of a Bi-2212 crystal for $\mu_0H = 1$ T and 5 T. In (a) the lines correspond to the direct thermal fluctuation effects in an homogeneous crystal. In (b) the effects of the $T_c$-inhomogeneities on the theoretical expressions have been taken into account, and also the average reduced temperature is used. Figures from Ref. 7.

The inhomogeneity effects on $\Delta\sigma_{ab}^e$ and $\Delta\tilde{\sigma}_{ab}^e$, have been estimated quantitatively in Ref. 7. For that, $\Delta\sigma_{ab}^e(\varepsilon,H)$, calculated through Eqs. (1) and (2) (and by using also the theoretical $\Delta\sigma_{ab}$ and $\Delta\tilde{\sigma}_{ab}$ for bilayered superconductors; see Ref. 9), was fitted to the experimental data with $\bar{T}_{c0}$ and $\Delta\bar{T}_{c0}$ as free parameters. As can be seen in Figs. 1 (b) and 2 (b), it is obtained an excellent and *simultaneous* agreement, the corresponding standard deviation for the critical temperature being $\Delta\bar{T}_{c0} = 0.6$ K. This value indicates that the stoichometric inhomogeneities, mainly of oxygen content, are quite small. But, in turn, these results of Ref. 7 clearly show that the presence of small $T_c$-inhomogeneities may dramatically affect the *measured* critical behaviour near $T_c$ of the HTSC. These results also provide an alternative explanation in terms of $T_c$-inhomogeneities to the in-plane magnetoconductivity anomalies observed by other

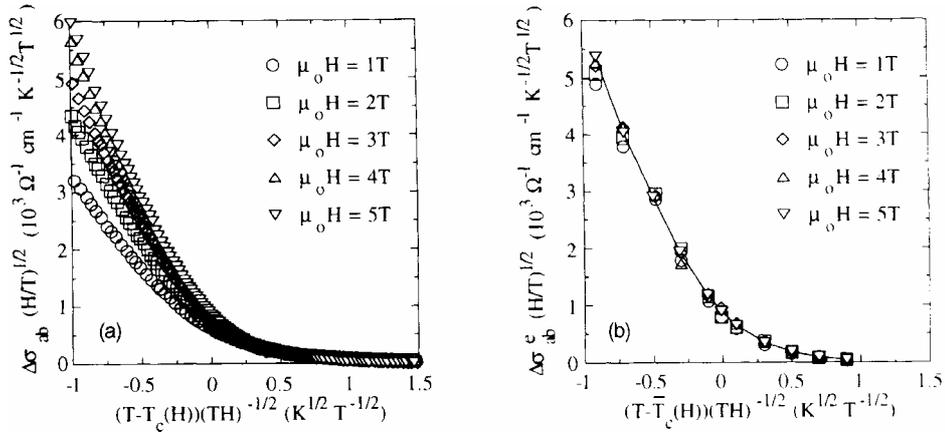

*Figure 2.* (a) Scaling of the excess conductivity for different magnetic fields measured in a Bi-2212 crystal. In (b) the effects of the $T_c$-inhomogeneities have been taken into account. The line through the 5 T data is a guide for the eyes. Figures from Ref. 7.

groups in Bi-2212 compounds and attributed by these authors to different *intrinsic* effects. [11]

## 2.2. SPLITTING OF THE RESISTIVE TRANSITION: INTRINSIC DOUBLE SUPERCONDUCTING TRANSITIONS VERSUS EXTRINSIC EFFECTS

The possibility of an intrinsic double superconducting transition in the HTSC was first open by some earlier heat capacity, $C_p$, measurements, which showed the presence in some samples of a double peak structure of the $C_p(T)$ curves near the average superconducting transition. [12] However, it was generally believed that these anomalies were probably an extrinsic effect, due for instance to the presence in these samples of stoichiometric inhomogeneities. The complexity of the HTSC chemistry, together with the strong influence of the oxygen content on their transition temperature, [13] made this simple explanation quite plausible. In fact, the extrinsic origin of the double transition was recently confirmed by heat capacity and magnetization measurements in $Y_1Ba_2Cu_3O_{7-\delta}$ (YBCO) samples with different oxygen contents. [14,15] However, in the last few years other groups, which have detected also double peak anomalies near $T_c$ in different HTSC by measuring different observables (heat capacity, magnetic susceptibility and electrical resistivity), have claimed that these effects are intrinsic and that they are related to central, although different from one group to another, characteristics of these materials. [16-18] For instance, in Ref. 16 the double peak structure observed in $C_p(T)$ near $T_c$ in the YBCO system is attributed to intrinsic $T_c$ variations associated with the oxygen configuration around the CuO chains of these compounds. In contrast, in Refs. 17 and 18 it is proposed that the anomalous double transition effect is a manifestation of the possible unconventional symmetry of the superconducting order parameter of these materials. The conclusions of Refs. 17 and 18 about the existence of an intrinsic double transition in the HTSC are based on the observation of a splitting of the bulk resistive transition of different compounds, the non observation of these effects in other $\rho$(T) measurements in HTSC being attributed to an insufficient experimental resolution. In fact, the presence of a double peak structure in the temperature derivative of $\rho$(T) was first stressed in Ref. 19, but these authors did not conclude about the origin of such an anomaly.

To prove the existence of an intrinsic double superconducting transition in HTSC we have performed in our laboratory high resolution measurements of the resistive transitions in different single crystal and polycrystal YBCO samples. [20] The temperature derivatives of these resistivity data were also analyzed. Here we are going to summarize some of these results which strongly suggest that the intrinsic resistive transition of the HTSC does not present any double transition anomaly, and that the double peak structure observed in $d\rho(T)/dT$ by some authors[16-19] is probably just an extrinsic effect associated, in some cases, with the presence in the samples of stoichiometric inhomogeneities (mainly, small oxygen content inhomogeneities). In other cases, this double peak structure could just be an experimental artefact due, for instance, to the great sensitiveness of the temperature derivative of $\rho$(T), near its maximum, to small electronic noise affecting the $\rho$(T) data points.

An example of the in-plane (parallel to the $CuO_2$ layers) resistivity obtained in a YBCO single crystal (sample noted YS25) is presented in the Fig. 3. For clarity, in Figs. 3(a) only about 5% of the measured data points has been plotted. These values of $\rho_{ab}(T)$ and of $d\rho_{ab}(T)/dT$ in the normal region well above the transition, in the temperature region where $\rho_{ab}(T)$ is a linear function of $T$, are typical of excellent YBCO crystals, although somewhat twinned. [21,22] Another indication of the high quality of this sample is provided by the exceptional sharpness of its resistive transition, whose half width may be defined by[21]

$$\left(\frac{d\rho_{ab}(T)}{dT}\right)_{T_{CI} \pm \Delta T_{CI}^{\pm}} = \frac{1}{2}\left(\frac{d\rho_{ab}(T)}{dT}\right)_{T_{CI}} \qquad (7)$$

where + or - correspond to, respectively, the upper and the lower half width and $T_{cI}$ is the temperature where $d\rho_{ab}(T)/dT$ around the transition has its maximum. As can be seen in Figs. 3(b) and 3(c), the total width of $d\rho_{ab}(T)/dT$ around $T_{cI}$ is, for this single crystal, less than 50 mK and 30 mK for, respectively, 3(b) and 3(c), the differences being associated with differences in the procedures used to obtain the temperature derivatives. In fact, these two figures also illustrate the crucial importance of the derivative procedure in analyzing with high temperature resolution the $\rho_{ab}(T)$ behaviour around $T_c$. In both cases, to obtain $d\rho_{ab}(T)/dT$ in each data point we use a polynomial determined by the best fit to its neighbour data points (together with the own data point). The use of a low degree polynomial or insufficient data points could introduce some spurious noise in $d\rho_{ab}(T)/dT$ which leads, in particular, to the onset of some kind of spurious structure around the $d\rho_{ab}(T)/dT$ maximum, the temperature region where the derivative is more sensitive to any irregularity. In Fig. 3(b), we have used a degree-three polynomial, that for each temperature was determined by the four nearest data points to such a temperature. This procedure, called algorithm A, leads to a quite regular derivative, *without any double peak structure*. As the spacing between data points is of the order of 5 mK, the resolution in temperature relative of this derivative is better than 20 mK. So, with a resolution better than 20 mK, the results summarized in Figs. 3(a) to 3(b) show the absence of a double resistive transition in YBCO single crystals. This temperature resolution must be compared with the temperature shift between the two peaks reported by different authors, [14-19] which was 50 mK or more. Complementary, the results of Fig. 3(c), have been obtained by using a degree one polynomial (called here algorithm B) and only with the two nearest data points to each given temperature. The corresponding structure of peaks around the $d\rho_{ab}(T)/dT$ maximum is clearly a spurious effect associated with this inadequate derivative procedure. Let us stress here that these results were also confirmed by the analysis of our previous measurements of the resistivity in the a-direction, $\rho_a(T)$, not affected by the presence of the CuO chains, in two untwinned YBCO crystals. [20] However, in these measurements the temperature shift between the data points was of the order of 20 mK and, therefore, the temperature resolution of $d\rho_a/dT$ around $T_{cI}$ was only of the order of 80 mK.

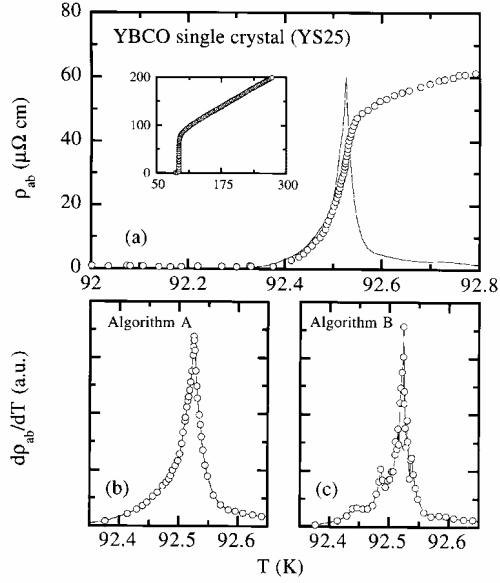

*Fig. 3.* (a) Temperature behaviour of the in-plane resistivity of one of the $Y_1Ba_2Cu_3O_{7-\delta}$ single crystals studied in this work (in this case, Ys7). In these two figures, for clarity only about 5% of measured data points are shown. (b) and (c) Temperature derivatives of the resistivity obtained by using the algorithms noted A and, respectively, B, and defined in the main text. Figure from Ref. 20.

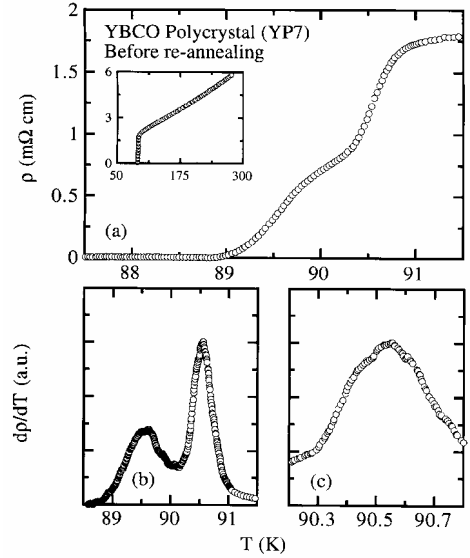

*Fig. 4.* (a) Temperature behaviour of the resistivity of one of the $Y_1Ba_2Cu_3O_{7-\delta}$ polycrystals studied in this work (in this case, sample YP7 before any re-annealing treatment). (b) Temperature derivative of the resistivity around the transition showing a two-peak structure typical of a two-phase sample. (c) Detail of $d\rho(T)/dT$ peak corresponding to the better oxygenated phase showing no sub-peak structure. Figure from Ref. 20.

As a complementary check of the possible existence of an anomalous resistive peak structure around $T_{cI}$ in the YBCO system, in Figs. 4 and 5 we present an example of the results obtained in ceramic YBCO samples before and after re-annealing. This example correspond to the sample noted YP7. As may be seen in Figs. 4, before re-oxygenation the $\rho(T)$ behaviour of this sample shows the typical kink of a two-phase sample, with two different $T_c$'s, their temperature shift being of the order of 1 K. This $\rho(T)$ behaviour is very similar to that of some of the YBCO samples studied in Ref. 16, and attributed by these authors to intrinsic effects associated with the oxygen arrangement around the CuO chains. However, such a double transition completely disappears after a new re-annealing, as shown in Fig. 5. This result clearly confirms that this double transition is just associated with the presence in the initially deficient oxygenated sample of small oxygen content inhomogeneities (less than 4%, the resolution of our x-ray analysis), at long length scales (i.e., at length scales much larger than the superconducting correlation length, even for temperatures relatively close to the transition) and uniformly distributed. A detail of the temperature derivative of the $\rho(T)$ part associated with the better oxygenated phase, with higher $T_c$, is presented in Fig. 4(c). We see here that within a temperature resolution of the order of 20 mK no sub-peak structure is observed. The absence of such an anomalous peak structure is also confirmed by the results presented in Fig. 5(b) for $d\rho(T)/dT$ of the re-annealed YP7 sample. We have also analyzed other ceramic YBCO samples, with different $T_c$'s (i.e., with different oxygen content). These analyses show that the absence of an anomalous peak structure of $d\rho(T)/dT$ is a general behaviour of the YBCO system, independently of its oxygen content, at least until $T_c$'s of the order of 80 K.

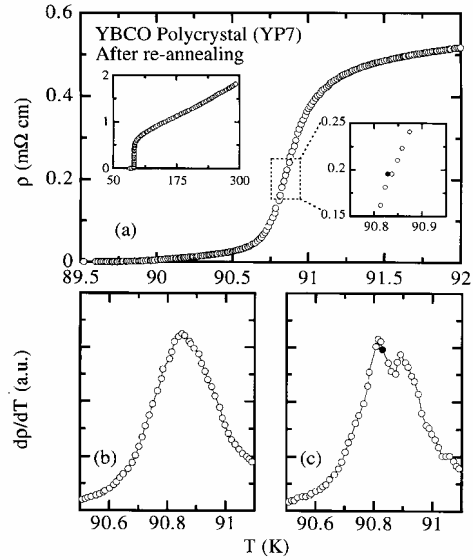

*Fig. 5.* (a) Temperature behaviour of the resistivity of the same polycrystalline sample as in Fig. 4 (sample YP7) but after a re-annealing treatment. In the scope in (a) we show an example of the defaults introduced artificially in the $\rho(T)$ curves to check the resolution of our temperature derivative procedure. This example consists in the shift in temperature (5 mK to lower temperatures) of a measured data point (open circle). The shifted data point is the solid circle. (b) and (c) Temperature derivatives of the resistivity obtained from the measured data points and, respectively, from the $\rho(T)$ curve resulting after the introduction of the default shown in the scope in (a). The solid circle here corresponds to the shifted point in the scope in (b). Figure from Ref. 20.

To check that the temperature derivative procedure introduced in Ref. 20 and based on the algorithm A, was able to detect the possible presence of small $\rho(T)$ anomalies around the transition, various types of artificial deformations were introduced in the experimental $\rho(T)$ curves. One of the deformations analyzed was the shift in temperature of just one of the $\rho(T)$ data points, mainly when it is located near $T_{cI}$. An example of such an artificial default, in this case introduced in the $\rho(T)$ curve of sample YP7 after re-annealing, may be seen in the inset of Fig. 5(a). The temperature shift between the measured data point (open circle) and the one artificially moved to the left (closed circle) is 5 mK. The resulting $d\rho/dT$ is presented in Fig. 5(c), which should be compared with the results of Fig. 5(b) for the non deformed $\rho(T)$ curve. This example clearly illustrates the dramatic influence on $d\rho(T)/dT$, mainly in the temperature region close to the derivative maximum, of small spurious shifts of the individual data points. In this example, we see that the spurious default in $\rho(T)$ originates a double peak structure around the transition that is very similar to that observed in Refs. 18 and 19 in YBCO samples, and attributed by these authors to intrinsic effects. These results of Ref. 20 strongly suggest, however, that these type of anomalies may probably be due to various types of spurious effects, as electronic noise affecting the resistivity or the temperature measurements. It is also very easy to check other general aspects of these spurious peaks of $d\rho(T)/dT$. For instance, the temperature shift between these peaks will directly depend on the total width of $d\rho(T)/dT$ around $T_{cI}$ (and this shift will be less than this total width). This may explain from one side why the peaks associated with the $\rho(T)$ noise will in general be much closer in temperature than those associated with the presence in the sample of various phases (compare the results of Figs. 4 and 5), and from the other side the differences observed in Ref. 18 between the temperature splitting of YBCO samples and the one of $Bi_2Sr_2CaCu_2O_{8+x}$.

We may conclude therefore that the results of Ref. 20 summarized here show the absence, with a resolution well to within 20 mK, of a double peak structure of $d\rho(T)/dT$. Note that the temperature shift between the two peaks reported by the different authors [12,14-19] was 50 mK or more, i.e., larger than our experimental resolution. Complementarily, we have analyzed the influence on $d\rho(T)/dT$ of various types of extrinsic effects including the presence in the samples of various stoichiometric phases as well as the presence of spurious noise effects in the resistivity or in the temperature measurements. The present results strongly suggest that the double peak structure of the resistive transition observed by various groups [16-19] in different high temperature copper oxide superconductors is an extrinsic effect. Moreover, as already stressed in Ref.19 at present does not exist any firm theoretical link between the presence of a possible double transition in the HTSC and the symmetry of its superconducting order parameter. [23] So, these different results on $d\rho(T)/dT$ do not allow to draw any conclusion about the wave pairing state in HTSC. [24]

## 3. $T_c$-inhomogeneities non-uniformly distributed

It is now well established that the presence of $T_c$-inhomogeneities at long length scales but *non-uniformly* distributed may generate various types of striking anomalies in the transport properties around the superconducting transition in HTSC, including the so-called anomalous peaks of the in-plane resistivity[25] the thermoelectric power[26] and the magnetoconductivity. [27,28] In addition to their intrinsic interest, these results may also provide an alternative explanation, in terms of temperature independent current redistributions associated with non uniformly distributed $T_c$-inhomogeneities, of the anomalous resistivity and magnetoresistivity peaks observed above the average superconducting transition by different groups in other LTSC[29-31] and HTSC[32-35] compounds and that are being attributed to different, and in some cases not well settled, intrinsic effects. [29-35] In the case of the thermopower, it is also possible to explain the anomalous $S(T)$ peaks observed in YBCO samples in terms of oxygen content inhomogeneities. [26] In this case, this explanation does not need any particular distribution of the inhomogeneities but it needs the simultaneous presence of two possible consequences of these inhomogeneities [26]: differences between the $T_c$'s of various sample domains and also differences between the sign of their corresponding thermopower. Both differences, in $T_c$ and in the sign of $S$, may be a consequence of small oxygen content differences in almost full doped YBCO samples, where in fact these $S(T)$ anomalies have been observed. [26,36] This provides a simple explanation of the anomalous $S(T)$ peaks observed by different groups and attributed in some cases to sophisticated but not well settled intrinsic mechanisms. [36-38] In this Section, we will summarize some of our results on the influence of $T_c$-inhomogeneities on the in-plane magnetoconductivity in inhomogeneous HTSC crystals. We will also briefly comment on the thermopower anomalous peaks in YBCO samples with small oxygen content inhomogeneities.

## 3.1. ANOMALOUS PEAKS OF THE IN-PLANE MAGNETORESISTIVITY AROUND $T_c$

An example of the in-plane magnetoresistivity peaks observed in Ref. 28 in a Y-123 crystal (sample Y16 of Ref. 28) before re-oxygenation are presented in Figs. 6(a) and (b). These anomalous peaks lead to a negative and anisotropic (i.e., depending on the orientation of $H$ with respect to the $CuO_2$ planes) in-plane magnetoresistivity excess which, for each magnetic field orientation, may be quantified through

$$\Delta\rho(T,H) \equiv \rho(T,H) - \rho(T,0) \qquad (8)$$

The data points in Figs. 7(a) and (b) correspond to the measured $\Delta\rho(T,H)$ for H perpendicular and, respectively, parallel to the ab planes. Only the $\Delta\rho(T,H)$ values corresponding to temperatures above the $\rho(T,0)$ peak have been represented. These important $\Delta\rho(T,H)$ values cannot be explained in terms of thermal fluctuations (see Ref. 3 and references therein). These results also show that, as may be expected from the $\rho(T,H)$ data, $\Delta\rho(T,H)$ is very anisotropic. In particular, for H parallel to the c direction the $\Delta\rho(H)_T$ saturation value is reached, at each temperature, for smaller field amplitudes than for H parallel to the ab planes. This behaviour of $\Delta\rho(T,H)$ is quite similar to that observed by other groups in other HTSC and LTSC compounds having $\rho(T,H)$ peaks near the superconducting transition and attributed by these authors to intrinsic effects[29,31-35] (compare, in particular, Figs. 7(a) and (b) with Fig. 2 of Ref. 34).

The experimental results for the Y16 sample after its reoxygenation are presented in Figs. 8(a) and (b) for $H$ applied parallel and, respectively, perpendicularly to the $CuO_2$ planes [see inset in Figs. 6(a) and (b)]. In this new oxygen annealing the crystal was again placed in a boat within a tubular furnace with $O_2$ flowing. The furnace was then heated up to 600 °C (at a rate of 100 °C/h), kept at this temperature for 2h, cooled to 400 °C in 1h and held at this temperature during four days. These latter processes were repeated three times. We see in Fig. 8(a) that the anomalous peak has completely disappeared from this $\rho_{ab}(T)$ curve. As, in addition, no structural changes were observed after these new annealings, these results confirm that the peak observed before is related to the presence in the sample of small (much less than 4% of the average oxygenation, the resolution of our x-ray diffraction measurements) oxygen-content inhomogeneities, that are strongly reduced by successive $O_2$ annealings. Let us note here that these results fully confirm the behaviour observed before in other YBCO crystals with resistivity peak anomalies [25] and, therefore, they provide new support to the explanation of the anomalous resistivity peak in terms of non uniformly distributed oxygen content inhomogeneities.

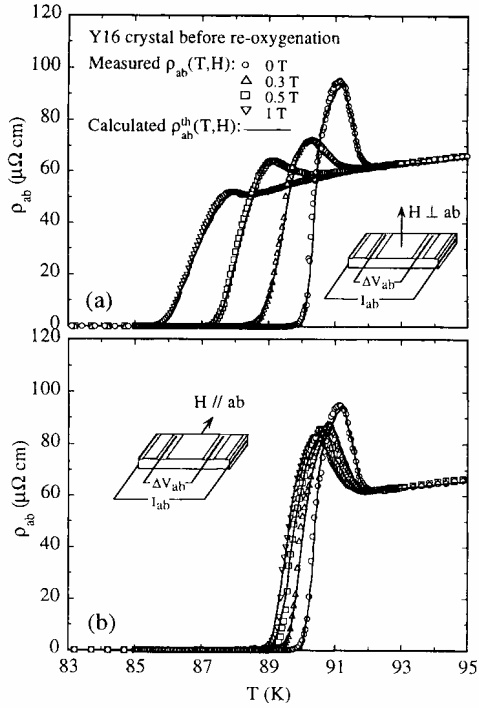

*Fig. 6.* In-plane magnetoresistivity versus temperature of the crystal Y16 before reoxygenation for different external magnetic fields with various amplitudes and orientations. In (a) the field was applied parallel to the crystallographic c-direction. In (b), it was applied parallel to the $CuO_2$ (ab) layers but still perpendicular to the injected current. The lines are the result of the simulations performed with the electrical resistor network represented in the inset of Fig. 8(b). Figures from Ref. 28.

These anomalous $\rho_{ab}(T,H)$ peaks have been explained in terms of non-uniformly distributed $T_c$-inhomogeneities in Ref. 28. Here we will summarize some of these results. Let us first stress here that these inhomogeneities are going to be small (of the order of two degrees or less, i.e., less than the 3% of the average $T_c$) and they may be due to small stoichiometric inhomogeneities (mainly of the oxygen content) extended over 10% or less of the sample volume. So, we are not able to directly determine these small local inhomogeneities. In addition, due to the oxygenation process, it is reasonable to expect that the best oxygenated parts of the crystal will be the edges of the sample domains. An example of an inhomogeneity distribution capable of generating the anomalous $\rho_{ab}(T,H)$ behaviour observed in the Y16 crystal is the one schematized in the inset of Fig. 8(a). In this figure, the shadowed domains at the upper edges of the crystal are better oxygenated and they have a higher $T_c$ than the rest of the crystal. The dimensions of each high- $T_c$ domain are $1/3 L_x$, $L_y$, $1/7 L_z$, where $L_x$, $L_y$ and $L_z$ are the sample's dimensions. To study how this inhomogeneity distribution affects the measured $\rho_{ab}(T,H)$ and generates the anomalous peak, we simulate the measurement through an equivalent electrical network. The geometry of the inhomogeneity distribution and the contact arrangement allows us to reduce the equivalent electrical network, in principle three dimensional, to the bi-dimensional one represented in the inset of Fig. 8(b). The resistances noted as $\widetilde{\widetilde{R}}_{ab}(T,H)$ and $\widetilde{\widetilde{R}}_c(T,H)$ correspond to the less oxygenated domains (with lower $T_c$), while $\widetilde{R}_{ab}(T,H)$ corresponds to the domains with higher $T_c$. Each resistance in the network is related to the corresponding resistivity in the crystal by

$$\widetilde{R}_{ab}(T,H) = \frac{L_x(N+1)}{L_z L_y N} \widetilde{\rho}_{ab}(T,H),  \qquad (9)$$

this relationship applying also to $\widetilde{\widetilde{R}}_{ab}(T,H)$ [with $\widetilde{\widetilde{\rho}}_{ab}(T,H)$], and by

$$\widetilde{\widetilde{R}}_c(T,H) = \frac{L_z(N+1)}{L_x L_y N} \widetilde{\widetilde{\rho}}_c(T,H) \quad . \tag{10}$$

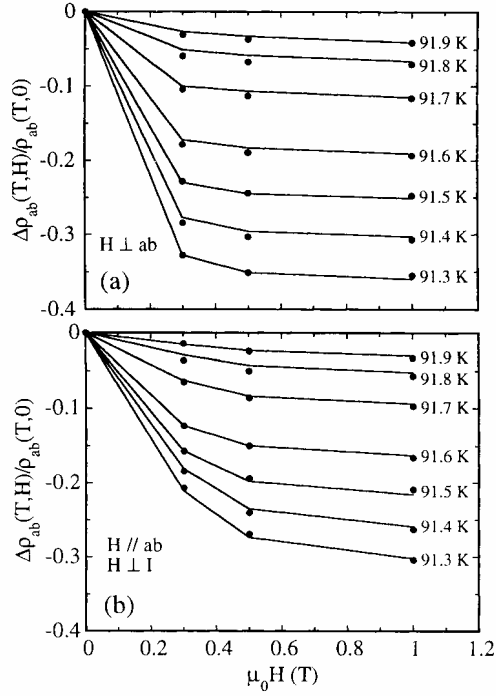

*Fig. 7.* Normalized magnetoresistivity excess of the crystal Y16 before reoxygenation at temperatures above but near the maximum of the anomalous $\rho(T,H)$ peak. (a) For $H$ applied parallel to the c-direction. (b) For $H$ applied parallel to the ab-planes and perpendicular to the injected current. The solid lines are the result of the simulation performed with the electrical resistor network represented in the inset of Fig. 8(b). Figures from Ref. 28.

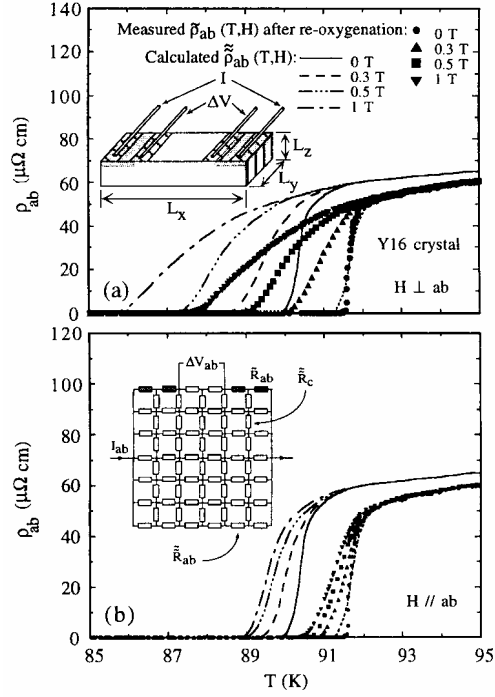

*Fig. 8.* In-plane magnetoresistivity versus temperature of the crystal Y16 after reoxygenation for different magnetic fields applied normally (a) and parallel (b) to the ab planes. The lines are the resistivities used in the electrical resistor network for the less oxygenated domains. Inset in (a): Schematic diagram of the $T_c$ inhomogeneities of the crystal Y16. The shadowed parts correspond to the highest $T_c$ domains and the dashed areas are the silver-coated electrical contacts. Inset in (b): Two dimensional electrical network for the sample schematized in (a). Figures from Ref. 28.

In these equations, $\widetilde{\widetilde{\rho}}_{ab}(T,H)$ and $\widetilde{\widetilde{\rho}}_c(T,H)$ correspond to the less oxygenated domains (with lower $T_c$), whereas $\widetilde{\rho}_{ab}(T,H)$ corresponds to the domains with higher $T_c$, and $N$x$N$ is the number of meshes of the network (6x6 in this case). For $\widetilde{\widetilde{\rho}}_{ab}(T,H)$, we used the profiles also shown in Figs. 8(a) and (b), that are typical of non-fully oxygenated Y-123. The resistivity in the c-direction is assumed to be one hundred times the in-plane resistivity. We assume also that $\widetilde{\rho}_{ab}(T,H)$ for $H$ parallel and perpendicular to the ab-planes may crudely be approximated by the resistivities measured after a new oxygen annealing, which are presented in Figs. 8 (a) and (b). However, to achieve the excellent agreement with the experimental results observed in Fig. 7, for $\widetilde{\rho}_{ab}(T,0)$ we have used the dotted curve represented in the same figure, that is slightly smoother than the experimental $\rho_{ab}(T,0)$. Such an excellent agreement is obtained in spite of the simplicity of our network, consisting only in 6x6 meshes and two different types of domains. [39]

A first example of the results of these calculations are the solid lines in Figs. 6 and 7. As it can be seen, the agreement between the experimental data and the simulation is excellent, the anomalous behaviour of the magnetoresistivity and of the magnetoresistivity excess being reproduced at a quantitative level for both orientations of the external magnetic field. In Fig. 9(a), it is represented the current distribution in the network at $T = 95$ K, a temperature well above the superconducting transition. This case corresponds, therefore, to the trivial situation in which the different sample domains with different oxygen content have almost the same (normal) resistivity, so the current lines are parallel to the ab plane and uniformly distributed and no anomaly is then observed. In contrast, the electrical current distribution shown in Fig. 9(b) corresponds to $T = 91.2$ K, the temperature at which the maximum of the resistivity peak occurs. At this temperature, the domains with higher $T_c$ are already superconducting and, therefore, the current density distribution is no longer uniform. There appears a higher current density in the top face of the crystal, where the voltage contacts are placed, giving rise to the anomalous voltage peak. Finally, in Fig. 9(c) it is represented the current distribution corresponding to $T = 91.2$ K in the presence of the external magnetic field of 1 T applied perpendicularly to the ab planes. The main effect of the magnetic field is to broaden the resistive transition, making the differences between the resistivities of the different domains much smaller than for H = 0. As a consequence, the current distribution is nearly uniform and the anomalous peak almost disappears from the effective (or measured) $\rho_{ab}(T,H)$ curves. Moreover, the broadening of the resistive transition is more pronounced for $H \perp$ ab than for $H$ // ab, and this is because the field amplitude needed to completely quench the anomalous peak is bigger for the latter field orientation.

As already noted in the introduction of this Section, anomalous magnetoresistivity peaks near $T_c$ very similar to the ones described here have been observed in some thin films and granular samples of different LTSC. [29] The simplicity of the chemical structure of these compounds makes quite improbable the presence of appreciable non-uniformly distributed compositional inhomogeneities. However, another source of $T_c$ inhomogeneities could be related to the well known low-dimensionality effects, which appear when the superconducting coherence length, $\xi(T)$, becomes of the order of or bigger than one of the sample's dimensions (thickness in the case of thin films, or grain diameter in the case of granular samples) [4]. Due to the relatively important coherence length amplitude of the LTSC [$\xi(T = 0\ \text{K}) \approx 1000$ Å], these low dimensionality effects may easily be present in the LTSC films. In addition, the thin film edges are never completely sharp, but instead they have an irregular thinner shape, and so having a higher $T_c$ than the rest of the film. When the temperature is well above any possible $T_c$ in the film, the electrical current used to measure the resistivity must be uniformly distributed. But at temperatures in which the film edges are superconducting and the rest of the film still remains in the normal state, the current lines must concentrate in the edges. This current redistribution may give rise to an increment in the signal detected by a voltmeter connected at the film edges and, in some cases, to the anomalous resistivity peak. A condition for that is simply that the edge thickness should be non uniform *along* the film. Otherwise, at the temperature at which the film edges are already superconducting a continuous superconducting path would connect both voltage terminals and no signal (and then no peak) would be detected. The anomaly so originated will be very dependent on the current used to perform the measurements. In fact, due to the smallness of the section of the highest $T_c$ edges, it is very easy to reach the critical current density by increasing the applied current, making the effect to disappear. Some of the different experimental results of Ref. 29 may satisfactorily be explained on the grounds of these simple ideas based on the presence of low dimensionality induced $T_c$ inhomogeneities, non uniformly distributed in the films. The anomalous $\rho(T,H)$ peaks observed in some granular LTSC samples[29] could also easily be explained in terms of low dimensionality effects, through the dependence of $T_c$ with the grain diameter, if a non-uniform distribution of grains having different diameters exists in the sample.

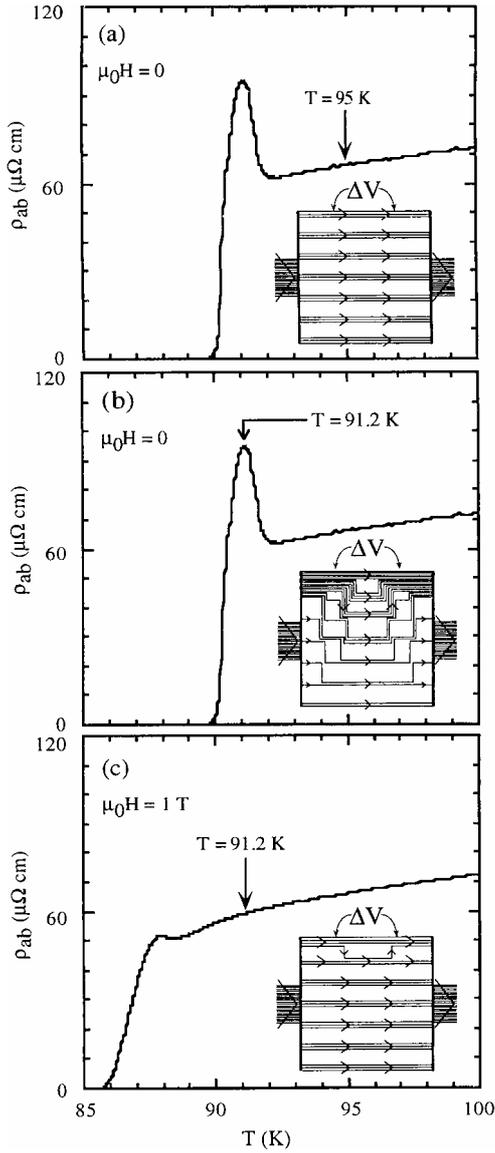

*Fig. 9.* Examples of the current redistributions originated in the electrical network corresponding to the crystal Y16 and represented in the inset of Fig. 8(b): (a) At $T = 95$ K, well above any superconducting transition in the sample. (b) At $T = 91.2$ K the temperature at which the anomalous peak has its maximum and the highest $T_c$ domains become superconducting. (c) At $T = 91.2$ K, with a magnetic field of 1 T parallel to the c-crystallographic direction. These striking differences in both the current distributions and in the measured $\rho(T,H)$ are associated just with differences of the temperature and magnetic field dependence of $\rho(T,H)$ in each sample domain. Figures from Ref. 28.

3.2. NEGATIVE IN-PLANE LONGITUDINAL VOLTAGES AROUND $T_c$.

The results summarized in the precedent subsections, clearly suggest that the magnetoresistivity anomalies observed in inhomogeneous HTSC could strongly depend on the type of spatial distributions of the $T_c$-inhomogeneities. In other words, different locations of the non-uniformly distributed $T_c$-inhomogeneities could originate very different behaviours of $\rho(T,H)$ around the average $T_c$. To illustrate this conclusion, here we are going to summarize some of our measurements of the in-plane longitudinal magnetoresistivity of $Tl_2Ba_2Ca_2Cu_3O_{10}$ (Tl-2223) crystals with stoichiometric inhomogeneities. In some of these crystals we have observed that the in-plane longitudinal voltage measured in presence of magnetic fields applied perpendicularly to the ab-layers, $V(H)$, is negative just below $T_c$. This anomaly may be explained in terms of $T_c$-inhomogeneities non-uniformly distributed in the sample *surface*.[27]

An example of $T_c$-inhomogeneities non-uniformly distributed in the sample surface are schematized in Fig. 10(a). This example was studied in more detail in Ref. 27. Here again, the shadowed parts correspond to the domains of higher $T_c$, the corresponding three-dimensional electrical network being represented in Fig 10(b). The resulting resistivity and current redistributions are represented in Figs. 10(c) and (d). The striking result is the appearance, just below the average superconducting transition and in some parts of the sample surface, of counter-currents with sign opposite to that of the injected electrical current. This may lead to the appearance of a negative effective resistivity just below the transition as it is illustrated in Fig. 10(c). These effects could explain the magnetoresistivity anomalies that we have recently observed in some inhomogeneous Tl-2223 crystals.[27] Besides, these negative surface currents associated with $T_c$-inhomogeneities non-uniformly distributed in the sample surface, could maybe contribute to explain some of the anomalous negative behaviour of other longitudinal and transversal transport properties observed around $T_c$ in LTSC[40] or HTSC[41] materials.

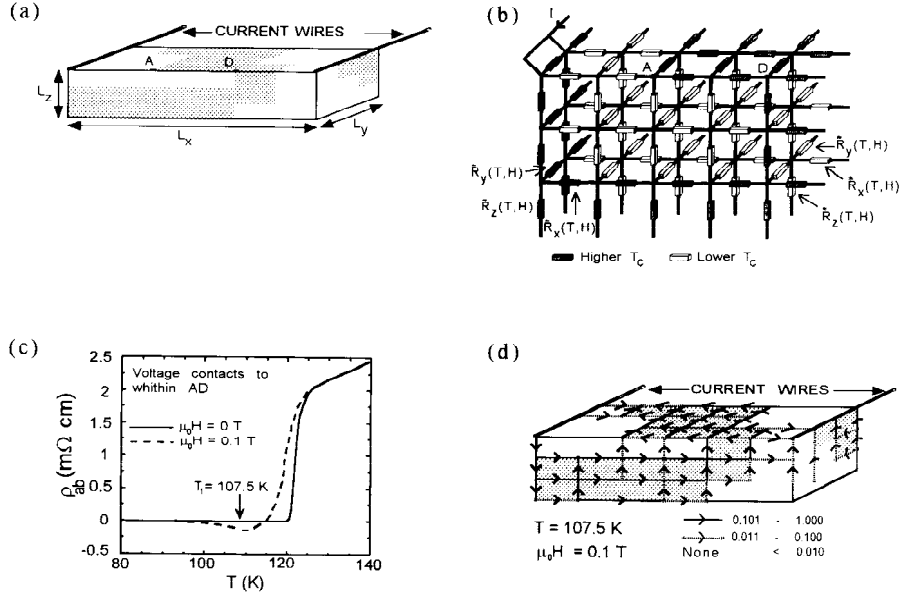

*Figure 10.* (a) Example of $T_c$-inhomogeneities non-uniformly distributed in the sample surface. (b) Corresponding three-dimensional resistor network. (c) Resulting in-plane effective resistivities for two applied magnetic fields. (d) Current density distribution at $T_1$ and $\mu_0 H = 0.1$ T. Figure from Ref. 27.

## 3.3. ANOMALOUS PEAKS OF THE THERMOPOWER AROUND $T_C$: A BRIEF COMPARISON WITH THE ELECTRICAL RESISTIVITY PEAKS.

Since the measurements of Cabeza and coworkers, it is now well established that the *intrinsic* critical behaviour of the thermopower, $S(T)$, near $T_c$ in copper-oxide superconductors is mainly driven by that of the electrical conductivity. In other words, the measurements in *homogeneous* HTSC show that their thermoelectric coefficient, $L(T)$, does not present any "sharp" critical divergence above the superconducting transition. [42] Instead, $L(T)$ in HTSC has around $T_C$ the logarithmic divergence earlier predicted by Maki. [43] Complementarily, since the measurements of Mosqueira and coworkers[26] it is now also well established that the presence of small oxygen content inhomogeneities in $Y_1Ba_2Cu_3O_{7-\delta}$ (Y-123) samples, *uniformly or non-uniformly distributed*, may also deeply affect the behaviour of $S(T)$ around the average superconducting transition. In fact this provides a simple, and now widely accepted, explanation, in terms of the oxygen content inhomogeneities, of the $S(T)$ anomalous peaks observed by different groups in Y-123 samples. [36,37] Until the results of Ref. 26, these anomalous $S(T)$ peaks were attributed by various groups to different intrinsic, but not well settled, effects. [37,38]

An example corresponding to the inhomogeneities sketched in Fig. 11(b) and (c), of the influence of the relative amplitude of the inhomogeneity domain on the effective (measured) thermopower peak is represented in Fig. 12. This example corresponds to non-uniformly distributed $T_c$-inhomogeneities. The details may be seen in Ref. 26. In this reference it was also studied the quenching of the anomalous $S(T,H)$ peak by a magnetic field, an effect observed by various authors[36] and which has remained unexplained until the results of Ref. 26. The details of the generation of the $S(T)$ peaks by the presence of oxygen content inhomogeneities in Y-123 crystals may be seen in Ref. 26. However, it may be useful to briefly compare here how these inhomogeneities originate the anomalous $S(T)$ or the $\rho(T)$ peaks. The *independent* peaks that near $T_c$ may present both observables in some $Y_1Ba_2Cu_3O_{7-\delta}$ samples may be understood in both cases in terms of oxygen content inhomogeneities at long length scales. However, in the case of $\rho(T)$, the anomalous peak may be explained by just taking into account only the associated $T_c$-inhomogeneities, which when they are non-uniformly distributed may lead near $T_c$ to strong electrical current density inhomogeneities in the sample.[25] In fact, these temperature-dependent current-density inhomogeneities are, independently of their origin, the only crucial ingredient for the appearance of these $\rho(T)$ anomalies. In contrast, in the case of $S(T)$ the possible heat current inhomogeneities (or, equivalently, the $\nabla T$ inhomogeneities) are practically irrelevant, and the appearance of the anomalous $S(T)$ peaks is, in general, just due to the addition of the different $S(T)$ associated with the various sample domains with different oxygenations. The appearance of a $S(T)$ peak will then need the simultaneous presence of the two possible consequences of the oxygen content inhomogeneities: differences between the $T_c$'s of the different sample domains but also differences between the sign of their corresponding thermopowers. As in the case of single phase $Y_1Ba_2Cu_3O_{7-\delta}$ samples the sign change of $S(T)$ appears for $\delta \approx 0.06$, the presence of an anomalous $S(T)$ peak will be possible only if one part of the sample is almost fully oxygenated. This is in contrast with the $\rho(T)$ peak, that only needs the presence of $T_c$-inhomogeneities. Another difference to be stressed here between both types of anomalous peaks concerns the inhomogeneity distribution. The appearance of a $\rho(T)$ peak will need a non-uniform distribution of the $T_c$-inhomogeneities, whereas the $S(T)$ peak may appear, as it has been shown by Mosqueira and coworkers in Ref. 26, even for uniformly distributed inhomogeneities. This explains why the $\rho(T)$ peak was observed only in crystal samples and not in polycrystals: In these last samples the $T_c$-inhomogeneities are in general uniformly distributed and they only broaden the resistive transition, without the generation of a peak. [6]

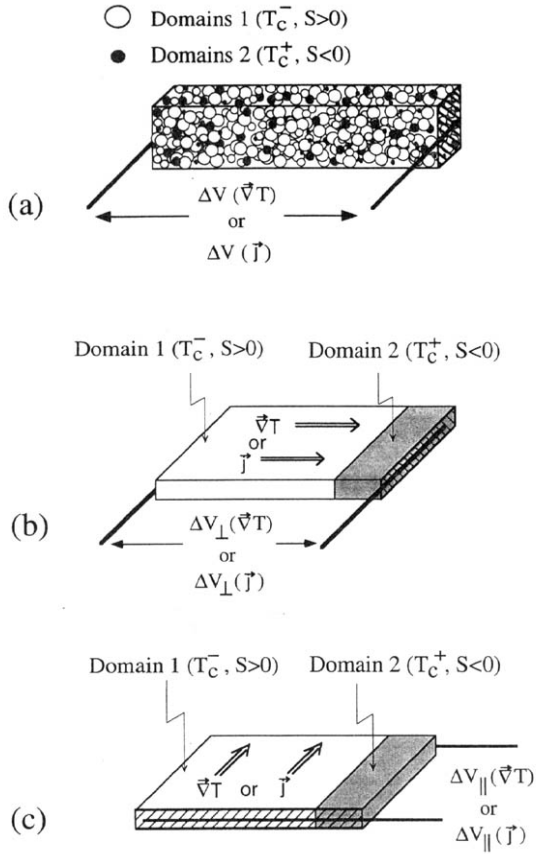

*Fig. 11.* (a) Schematic diagram of an example of uniformly distributed oxygen content inhomogeneities in polycrystalline samples. The better oxygenated domains (with $\delta \approx 0$) correspond to the smaller grains while the largest domains have a slight deficiency in oxygen content ($\delta \approx 0.1$). (b) and (c) Schematic diagrams of an example of non-uniformly distributed inhomogeneities in the oxygen content in single crystals for two different leads arrangements. The shadowed domains are well oxygenated ($\delta \approx 0$) while the white domains have $\delta \approx 0.1$. Figure from Ref. 26.

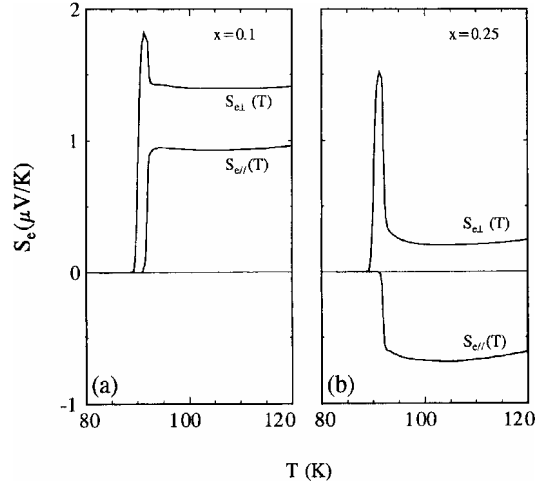

*Fig. 12.* An example, corresponding to situations sketched in Figs. 11(b) and (c), of the influence of the relative magnitude of the inhomogeneity domains on the effective thermopower in the case of non-uniformly distributed oxygen content inhomogeneities. In (a), x = 0.1, whereas in (b), x = 0.25. Figure from Ref. 26.

# 4. Two examples of other inhomogeneity effects: How the crossing point of the magnetization and the full critical behaviour of the paraconductivity may be affected by the presence of inhomogeneities.

For completeness, we will summarize here two examples of other inhomogeneity effects. The first example will concern the thermal fluctuation effects of vortices below $T_c$ in presence of stoichiometric and structural inhomogeneities. The second example will concern again the in-plane resistivity but this time very close to $T_c$ and in presence of both types of $T_c$-inhomogeneities at long length scales: uniformly and non-uniformly distributed. In this last case both type of $T_c$-inhomogeneities are going to be analyzed separately, but they may simultaneously affect $\rho_{ab}(T)$ very close to $T_c$.

## 4.1. THE CROSSING POINT OF THE MAGNETIZATION IN HIGHLY ANISOTROPIC HTSC IN PRESENCE OF STRUCTURAL AND STOICHIOMETRIC INHOMOGENEITIES.

Although this review is centered on the effects of the $T_c$-inhomogeneities on the in-plane transport properties, it will be useful to note here that the presence in the samples of structural inhomogeneities, always at long length scales, may also deeply affect any observable and, in particular, the *measured* behaviour of $\rho(T,H)$ and of $S(T,H)$ around $T_c$. These structural inhomogeneity effects at long length scales, as those that exist in granular and ceramic HTSC, on $\rho(T,H)$ and on $S(T,H)$ around $T_c$ have been earlier analyzed in our group, and a procedure to separate them from the intrinsic fluctuation effects was also proposed. [42,44,45] The interest of these anomalies of the structural inhomogeneity effects on $\rho(T,H)$ is enhanced by the fact that they are also directly related to the critical current densities in granular and ceramic HTSC. [46] However, these analyses are out of the scope of our present review. Some of the main aspects of these effects and of their interplay with the thermal fluctuations may be seen in Refs. 42, 44 and 45. Here, to illustrate the influence of these structural inhomogeneities al long length scales we are going to just summarize briefly some of our recent results on the in-plane magnetization, $M_{ab}(T,H)$, around $T_c$ in highly anisotropic HTSC with small structural and stoichiometric inhomogeneities. Note also that until now we have reviewed in this paper some inhomogeneity effects on transport properties and its interplay with thermal fluctuations of Cooper pairs *above $T_c$*. In contrast, this new example concerns a static parameter and the interplay between stoichiometric and, mainly, structural inhomogeneities with thermal fluctuations of magnetic vortices *below $T_c$*.

The so-called "crossing point" of the excess magnetization versus temperature curves at a given magnetic field amplitude (with $H$ applied perpendicularly to the $CuO_2$ layers), $\Delta M(T)_H$, probably provides one of the best and easiest scenarios to check the interplay between structural and stoichiometric inhomogeneities at long length scales and thermal fluctuation effects in HTSC. This crossing point of the $\Delta M(T)_H$ curves was first observed experimentally by Kadowaki[47] and, independently, by Kes and coworkers. [48] In the case of highly anisotropic HTSC and for high magnetic field amplitudes (i.e., for $H \gtrsim (1/7)H_{c2}(T^*)$, where $H_{c2}$ is the upper critical field at the crossing point temperature, $T^*$), these fluctuation effects are attributed to the creation and anihilation of quasi-two dimensional (pancake) vortices. [49] In this case, the Ginzburg-Landau model in the so-called lowest-Landau-level approximation (GL-LLL) predicts that the crossing point coordinates are related by[49]

$$-\Delta M^*_{ab} = \frac{k_B T^*}{\phi_0 s^v_e}, \qquad (11)$$

where $k_B$ is the Boltzmann constant, $s^v_e$ is an effective periodicity length which takes into account the possible multilayering effects on the vortex fluctuations and $\phi_0 = h/2e$ is the flux quantum (*h* is the Planck constant and *e* is the electron charge). The importance of Eq. (11) is enhanced by the fact that it relates directly the effective periodicity length, $s_e^\gamma$, a microscopic parameter which in multilayered HTSC may depend on the Josephson and on the magnetic couplings between adjacent superconducting layers, to two directly measurable macroscopic observables, $\Delta M_{ab}^*$ and $T^*$, and that without any dependence on $T_{c0}$, the mean-field critical temperature, which is never directly accessible. This theoretical result has lead, therefore, to much experimental activity in the last years. [50] However, all the $\Delta M^*/T^*$ data published until now in polycrystalline or in single crystalline HTSC strongly disagree, in both the amplitude and the s dependence, with Eq. (11). In particular, in most of the experiments the measured $\Delta M_{ab}^*/T^*$ leads to an effective periodicity length, $s_e^\gamma$, larger than s, in contradiction with the theoretical predictions. [49,51] An example of the disagreement between the measured crossing point coordinates and Eq. (11) may be seen in Fig. 13(a). This example correspond to a TlPb-1212 crystal. [52] The presence of strong stoichiometric inhomogeneities, which will appreciably reduce the superconducting fraction, has been discarded in most of the studied samples by independent measurements (x-ray and neutron diffraction, in particular). [50] Therefore, until now most of the authors propose that these $\Delta M_{ab}^*/T^*$ data are intrinsic and that the BLK and the TXBLS approaches do not explain, even at a qualitative level, the crossing points observed in highly anisotropic HTSC. [50]

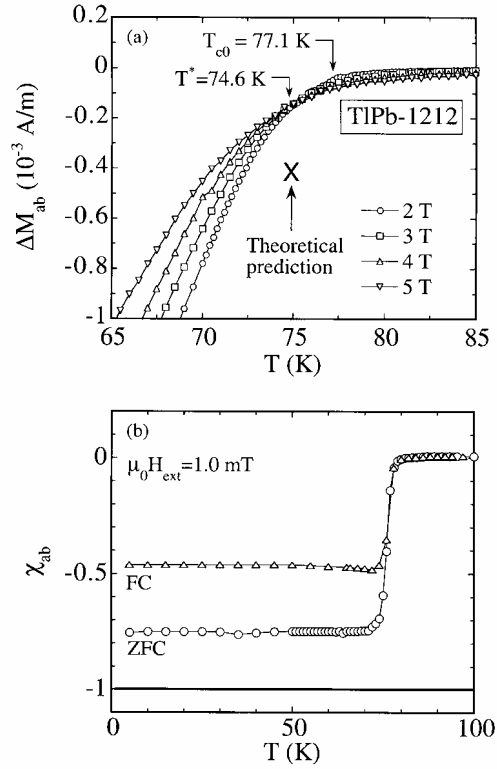

*Fig. 13.* (a) Measured in-plane excess magnetization vs. temperature, at different constant magnetic fields applied normally to the ab-planes, of a TlPb-1212 single crystal. (b) Corresponding in-plane field-cooled and zero-field-cooled susceptibilities. The main error source in these last observables is the demagnetization factor uncertainties.

It has been proposed recently, however, that the strong disagreement between the experimental data and the crossing point coordinates predicted by the theory of Tešanovic and coworkers[49] could be resolved by taking into account all the possible non intrinsic effects on the magnetization. [52] These non intrinsic effects will be associated with structural and stoichiometric inhomogeneities, at different length scales and amplitudes, and not only with those due to the presence of strong stoichiometric inhomogeneities at long length scales (i.e., at length scales much larger than the superconducting coherence lengths, which are those easily observable with conventional x-ray and neutron diffraction techniques). This conclusion was strongly supported by simultaneous measurements of the crossing point in the high-magnetic-field limit [$H \bigcirc H_{c2}(T^*); H \prec H_0$] and of the field-cooled susceptibility (the so called Meissner fraction), $\chi_{ab}^{FC}$, in different single crystals of various highly anisotropic HTSC families with different values of *N* and *s*. An example of the field-cooled (FC) and zero-field-cooled susceptibilities may be seen in Fig. 13(b). This example corresponds to the same sample than in Fig. 13(a). It may be easily concluded from Figs. 13(a) and (b) that the difference between the measured $\chi_{ab}^{FC}(T^*)$ (already corrected for demagnetization effects) and the susceptibility of an ideal superconductor is, within the experimental uncertainties, the same as the difference between the measured $\Delta M_{ab}^*$ the in-plane excess magnetization predicted by Eq. (11), with $s_e^\gamma = s$. Similar measurements have been done in different HTSC with different values of *N* and *s*, in Ref. 52. These results demonstrate experimentally that in highly anisotropic HTSC crystals $\Delta M^*_{ab}/\text{-}\chi^{FC}_{ab}(T^*)\text{-}$ verifies, within the experimental uncertainties, Eq. (11), with $s^v_e = s$, independently of $N$. Complementarily, these results show that in spite of the fact that $\Delta M^*_{ab}$ and $\chi^{FC}_{ab}(T^*)$ are measured under very different magnetic field amplitudes [$H \bigcirc H_{c2}(T^*)$ and, respectively, $H \leq H_{c1}(T^*)$, the lower critical magnetic field at $T^*$], the non intrinsic effects on both observables, associated with stoichiometric and structural inhomogeneities at different length scales, are the same within the experimental uncertainties. $\Delta M^*_{ab}/\text{-}\chi^{FC}_{ab}(T^*)\text{-}$ is, therefore, the intrinsic excess magnetization coordinate of the crossing point.

The above results on the crossing point versus the Meissner fraction, obtained on relatively good single crystals and for magnetic fields applied perpendicularly to the ab planes have been extended recently to granular samples in Ref. 53. In this paper, not only the inhomogeneities but also the random orientation effects have been separated from the intrinsic vortex fluctuation effects. Let us finally note here that the correction of the inhomogeneity effects on $\Delta M(T)_H$ through $\chi^{FC}_{ab}(T)$ is just an approximation which does not apply necessarily to all samples. In fact, we have experimentally observed the failure of such a correction for some of the very inhomogeneous samples studied, with quite low $\text{-}\chi^{FC}_{ab}(T^*)\text{-}$ values (let us say, with $\text{-}\chi^{FC}_{ab}(T^*)\text{-} \bigcirc 0.3$).

## 4.2. FULL CRITICAL BEHAVIOUR OF THE PARACONDUCTIVITY VERSUS $T_C$-INHOMOGENEITIES UNIFORMLY AND NON-UNIFORMLY DISTRIBUTED.

Due to the high amplitude of their thermal fluctuations, it was earlier recognized that the HTSC could be excellent candidates to experimentally penetrate in the so-called full critical region. [3,54] In that region, the amplitudes of the thermal fluctuation effects are expected to be even bigger than the amplitudes of each observable in absence of fluctuations and, therefore, these effects cannot be any more considered as a small perturbation of the mean field like behaviour. Therefore, their corresponding critical exponents are expected to be different from those of the so-called mean-field-like region. For instance, in the case of the paraconductivity, the 3D-XY theory for full critical fluctuations predicts a critical exponent for the in-plane paraconductivity equal to $-(2/3)(z-1)$, where z is the so-called dynamic critical exponent. [3,54,55] If, in addition, z=3/2, as predicted by the so-called E-model dynamics[56] (and which correspond to, for instance, the superfluid transition in the 4He liquid), then the critical exponent of the paraconductivity in the full critical region becomes -1/3, instead of -1/2 for the 3D mean-field region.

To our knowledge, the first experimental attempt to observe the full critical behaviour of the paraconductivity in HTSC was published in Refs. 56 and 57. In fact, an apparent critical exponent of -1/3 was observed for reduced temperatures below $\varepsilon \equiv (T-T_c)/T_c \bigcirc 10^{-2}$, in excellent qualitative agreement with the estimated Ginzburg-Levanyuk reduced temperature, $\varepsilon_{LG}$, which corresponds to the limit (closer to $T_c$) of the mean field like region estimated in these compounds.[3] However, as the samples used in these experiments presented a quite wide resistive transition (the temperature width of d$\rho$/d$T$, $\Delta T_C$, was of the order of 0.5 K or bigger), these results were not conclusive: as already stressed in these papers, this apparent full critical behaviour of the paraconductivity could as well be due to uniformly distributed $T_C$-inhomogeneities, associated for instance with relatively small oxygen content inhomogeneities. Moreover, a first analysis of the influence of the choice of $T_C$ on the paraconductivity behaviour and of the uncertainties on the critical exponents associated with the uncertainties in the precise location of $T_C$, was already presented in these references, and later in Ref. 6 (see also the note in Ref. 58). In addition, the polycrystallinity of the samples used in these experiments prevented the use of the paraconductivity *amplitudes* as a further check of the intrinsic full critical and mean field like behaviours. So, it was concluded in these earlier papers that the measurements in samples having a relative resistive width, $\Delta T_C/T_C$, bigger than $\varepsilon_{LG}$, will not allow any quantitative conclusion on the full critical region: For these samples, the apparent full critical behaviour could just be associated with the extrinsic roundings due to $T_C$-inhomogeneities. [59]

Further attempts to observe experimentally the full critical behaviour of the para-conductivity were done by measuring the in-plane resistivity in apparently high quality single crystals, with very sharp resistive transitions, $\Delta T_C$ being of the order of 0.1 K. An earlier example of these attempts may be seen in Ref. 60, where a critical exponent of $-1/3$ was observed for $\varepsilon \leq 10^{-2}$. In addition, the corresponding absolute amplitude agrees with that expected by scaling the mean-field amplitude (obtained by taking into account the presence of two Josephson coupled $CuO_2$ layers per periodicity length) to the full critical region, and also such a behaviour is consistent with the in-plane fluctuation induced diamagnetism measured in the same untwinned crystals. Also, the temperature used as $T_{c0}$, $T_{cI}$, agrees at a quantitative level with the critical temperature estimated through $\Delta\chi_c$, the excess diamagnetism for $H$ in the weak amplitude limit and applied parallel to the ab layers, that is not appreciably fluctuations. [60,61] (In fact, probably the best possible estimation of the mean field critical temperature, never accessible directly, is through $\chi_C(T,H \to 0)$). Therefore, probably the results of Ref. 60 provide the more convincing, still at present, [59] experiment probing the full critical behaviour of the para-conductivity in any HTSC (see later).

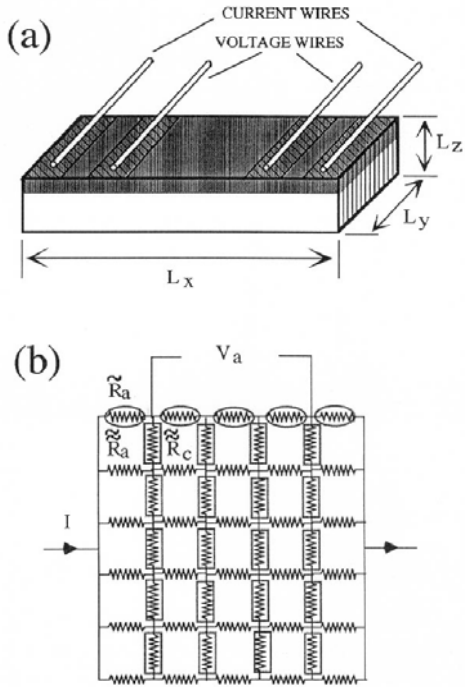

*Fig. 14.* (a) Schematic diagram of the sample YS3. For this crystal, the x direction in the figure corresponds to the crystallographic a direction. The shadowed parts correspond to the well-oxygenated domains and the dashed areas are the silver-coated electrical contacts. (b) Two-dimensional electrical-circuit model for the sample schematizaed in (a). Figure from reference 25.

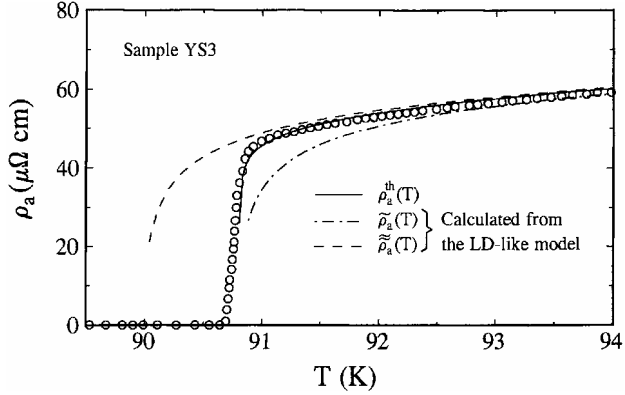

*Fig. 15.* Circles correspond to the measured resistivity in the a direction for sample YS3 around the transition. These data have been taken from Fig. 2(c) of Ref. 21. The dashed and dotted-dashed lines correspond to the resistivities calculated from the Lawrence-Doniach-like model by using different critical temperatures. The solid line is the result of the simulation performed with the electrical-circuit model of Fig. 14(b). Figure from reference 25.

However, as already stressed in that paper and later in Ref. 25, even in the highest quality samples, with very sharp resistive transitions, we cannot exclude the possible influence on the $\rho_a(T)$ behaviour of small $T_c$-inhomogeneities, in this case *non-uniformly distributed* in the crystals, associated with, for instance, a somewhat better oxygenation of the crystals surface than the inside of the crystals. An example of these inhomogeneities is shown in Fig. 14. As first shown in Ref. 25, this type of inhomogeneity may deform the $\rho_a(T)$ curve above but very close to the transition, at temperature distances to the transition of the order of the transition temperature difference, $\Delta T_c$, between the surface and the inside of the crystals, but *without* broadening the measured resistive transition.

An example of such a deformation is presented in Fig. 15. The data points in this figure correspond to $\rho_a(T)$, the resistivity in the a direction measured by Pomar and coworkers in an untwinned $Y_1Ba_2Cu_3O_{7-\delta}$ single crystal. [21] The solid line in this figure was generated with the electrical circuit model of Fig. 14(b), with $\tilde{R}_a$, $\tilde{\tilde{R}}_a$ and $\tilde{\tilde{R}}_c$ given by Eqs. (9) and (10). In these equations, $\tilde{\rho}_a(T)$ and $\tilde{\tilde{\rho}}_a(T)$ are obtained through, [21]

$$1/\rho_a(T) = \Delta\sigma_a(T) + 1/\rho_{aB}(T), \qquad (12)$$

where for $\rho_{aB}(T)$, the background resistivity, we use the values measured in Ref. 21. Although the background resistivity of each domain would be different, the use in our approximation of the same background resistivity for both domains does not have any qualitative influence in the results. For the paraconductivity, $\Delta\sigma_a(T)$, we use the Lawrence-Doniach (LD)-like expression, with the corresponding parameters also obtained, in the so-called mean-field region, in Ref. 21. The only difference between $\bar{\rho}_a(T)$ and $\tilde{\bar{\rho}}_a(T)$ is, therefore, their critical temperatures that are, respectively, 90.8 K and 90 K. As can be seen in Fig. 15, the agreement between $\rho_a^{th}(T)$ and the measured data is excellent, even very close to the measured transition, in a temperature region that could be affected by (full) *critical* order parameter fluctuations (OPF). [21] When combined with the analysis made in Ref. 21, the present results show that the resistivity rounding very close to the transition (for reduced temperatures of $\varepsilon = 10^{-2}$ or less) could be explained by the presence of intrinsic (full) critical OPF or, alternatively, by mean-field-like OPF (as those calculated on the grounds of the LD-like approaches) plus small $T_c$ inhomogeneities, associated with very small oxygen content inhomogeneities. However, let us stress here that $\rho_a(T)$ in the MFR (i.e., for $\varepsilon \gtrsim 10^{-2}$) will not be affected by these possible $T_c$ inhomogeneities. Let us also stress here that the $T_c$-inhomogeneity distribution of Fig. 14(a) is just an example, and that other distributions also will produce a $\rho_a(T)$ deformation. This can be, for instance, the case of a central part of the crystal with lower $T_c$ involved by a part (better oxygenated) with higher $T_c$.

The results summarized here just intend to show how crucial is to carefully check the possible extrinsic effects associated with structural and stoichiometric inhomogeneities in analyzing the critical behaviour of any observable very close to $T_c$ in HTSC. In fact, as noted already in the Introduction, the complicate chemistry of these materials, together with the strong sensitiveness of their $T_c$ to the stoichiometry appears, mainly in analyzing their critical behaviour very close to $T_c$, as the "counterpoint" to the high amplitude of their thermal fluctuations. Will these difficulties associated with inhomogeneities prevent any quantitative conclusion on the full critical region in HTSC? Although the results summarized here clearly show that very small inhomogeneities, almost undetectable by using conventional x-ray or neutron diffraction techniques, will suffice to strongly deform the behaviour of any observable very close to $T_c$, the answer is indeed not. In fact, systematic and reproducible data for different samples, for both *the critical exponent and the amplitude*, will strongly suggest an intrinsic behaviour. This was, in fact, the case of the results of Ref. 60: the data obtained in two different Y-123 crystals agree each other at a quantitative level, well to within the experimental uncertainties. As noted before, this provides, therefore, a quite convincing probe of the full critical behaviour, and also of the mean field like behaviour at bigger reduced critical temperatures!, of the paraconductivity in Y-123 crystals. Mainly in the case of the mean-field-like region, these conclusions were reinforced by independent paraconductivity measurements in other Y-123 untwinned crystals[62] and also by a recent comparison with other observables. [63,64] New reliable measurements of both the amplitude and the $\varepsilon$-behaviour of the paraconductivity very close to $T_c$ ($\varepsilon < 10^{-2}$) in other high quality HTSC crystals will be, however, desirable.

## 5. Conclusions

The examples reviewed here show that the presence of small stoichiometric and structural inhomogeneities may deeply affect, even when they have characteristic lengths much bigger than those of the superconductivity, the behaviour of any observable around $T_c$ in HTSC. These extrinsic effects will arise simultaneously with the intrinsic thermal fluctuation effects. So, in analyzing the measurements of these last effects, it will be crucial to detect and to separate them from those associated with inhomogeneities. But, in addition, these inhomogeneity effects may concern other important fundamental and practical aspects of the HTSC. For instance, the results about $\rho(T,H)$ in presence of $T_c$-inhomogeneities clearly suggest that the local current redistributions that these inhomogeneities introduce may also deeply affect the critical current in practical HTSC. Moreover, there is considerable room for further work, in particular, to understand how the "intrinsic" inhomogeneous current redistribution associated with the Meissner effect, may affect many properties around $T_c(H)$, or to extend the present results to $T_c$-inhomogeneities with short characteristic lengths (when compared with the superconducting characteristic lengths).

## 6. Acknowledgements

This work was supported by the Spanish CICYT, under grants Nº MAT 95-0279 and MAT 98-0371, the European Community Grant No. CHRX-CT93-0325, and by a grant from Union Fenosa, Spain, No.98-0666. CC acknowledges financial support from the Fundación Ramón Areces, Spain.